\documentclass[conference]{IEEEtran}
\IEEEoverridecommandlockouts
\usepackage{cite}
\usepackage{amsmath,amssymb,amsfonts}
\usepackage{algorithmic}
\usepackage{graphicx}
\usepackage{textcomp}
\usepackage{xcolor}
\usepackage{booktabs}
\usepackage{tabularx}
\usepackage[most]{tcolorbox}
\usepackage{CJKutf8}
\usepackage{ragged2e}
\usepackage{booktabs}
\usepackage{multirow}
\usepackage{colortbl}
\usepackage[table]{xcolor}
\usepackage{balance}
\usepackage{url}
\usepackage[hidelinks]{hyperref}

\def\BibTeX{{\rm B\kern-.05em{\sc i\kern-.025em b}\kern-.08em
    T\kern-.1667em\lower.7ex\hbox{E}\kern-.125emX}}
\begin{document}

\title{What Do Deepfake Audio Detectors Hear? From Self-Supervised Models to Audio LLMs\\
}

\author{
\IEEEauthorblockN{
Wanying Ge, Xin Wang, Junichi Yamagishi
}
\IEEEauthorblockA{
National Institute of Informatics \\
Chiyoda-ku, Tokyo 101-8430, Japan \\
\texttt{\{gewanying,wangxin,jyamagis\}@nii.ac.jp}
}
}

\maketitle

\begin{abstract}
Real-world audio recordings for deepfake detection rarely consist solely of speech; they often are mixtures of speech and background events that themselves may also be synthetic. When processing these mixtures, it remains unclear whether detectors attend to the speech or the background event, and where their spectral evidence lies. We study this on the CompSpoof V2 database, whose mixtures pair independently real-or-fake speech and event components. Our results show that all systems detect fake events more accurately than fake speech. However, this performance gap is largely driven by poor generalization to an unseen fake speech domain, rather than a true difference in difficulty. Furthermore, sub-band probing reveals that the front-end encoder dictates where a countermeasure focuses its frequency evidence. To enable reliable deployment in real-world scenarios, future countermeasures should address out-of-domain generalization failures and carefully select front-end encoders aligned with their target acoustic evidence. Our source code and checkpoints are released at \url{https://github.com/nii-yamagishilab/AntiDeepfake-AudioLLM}.
\end{abstract}


\vspace{-3.2pt}
\section{Introduction}

A deepfake speech countermeasure (CM) is, almost by construction, assumed to listen to the speech. Its front-end is pre-trained on speech recordings~\cite{baevski2020_wav2vec,wavlm}; its training data is curated to expose synthesis artefacts in the voice~\cite{wang2020_asvspoof2019,yi2023_add2023}. While the best detectors report near-perfect performance on speech deepfake corpora~\cite{jung2022aasist,tak2022automatic}, the same detectors degrade sharply once a recording is no longer isolated speech~\cite{muller22_interspeech,li2026howwell}: overlaying even a short non-speech segment onto synthetic speech can push a state-of-the-art detector close to chance level. This blocks deployment, because real recordings are rarely clean speech~\cite{yi2024scenefake,rahman2024sonics,zang2024singfake}. 

Though non-speech components are usually treated as interference~\cite{chandra2026dfe}, this very component, when synthetic, carries its own tell-tale artefacts~\cite{zhang2026esdd2overview,ouajdi2024detection,yi2024scenefake}. It is not known which component an audio detector actually follows~\cite{whatdoes}, nor which part of the spectrum carries that artefact~\cite{wang2022investigating}. We study this question on the CompSpoof V2 database~\cite{zhang2026esdd2}, in which the speech and the environmental sound components can each be real or fake independently. We pose two research questions: 
\begin{enumerate}
    \item RQ1 (component sensitivity). When speech and an event are mixed, which component does a detector catch more readily?
    \item RQ2 (sub-band dependence). Which frequency sub-bands carry the detector's evidence? 
\end{enumerate}

By scoring each fake component in isolation against a genuine mixture, and by muting one sub-band at a time, we find that fake events are detected more accurately than fake speech by every CM we test, but further source-level decomposition shows the performance gap is largely caused by poor generalisation towards a single unseen domain rather than a true difficulty gap. Sub-band probing further shows that the front-end encoder, rather than the classifier backbone, determines where a detector's evidence concentrates in frequency.

The remainder of the paper is organised as follows. Section~\ref{sec:cm} describes the countermeasures used in our work. Section~\ref{sec:exp_setting} details the data, metric, training detail, and sub-band probing protocol. Component sensitivity (RQ1) is analysed in Section~\ref{sec:rq1} and sub-band dependence (RQ2) in Section~\ref{sec:rq2}. Discussion and conclusions are reported in Section~\ref{sec:conclusion}.

\section{CM systems}
\label{sec:cm}

In this section, we describe the evaluated detectors. We compare two CM families (Figure~\ref{fig:cms}) to capture the paradigm shift from specialised speech detector to general audio understanding. The first is the standard recipe: a pre-trained encoder followed by a classification head that maps representations to a binary \texttt{\{real, fake\}} decision~\cite{wang2022investigating,tak2022automatic}. The second is the increasingly popular audio large language model (ALLM): this architecture feeds features from a pre-trained audio encoder into a transformer-based LLM to jointly process audio and text embeddings for a task-specific response~\cite{xie2026ftgrpo,li2025dfallm,gu2025allm4add}.

\subsection{Standard CM}
\label{sec:encoder-cm}

The input to a CM is the raw waveform. An encoder then extracts a sequence of frame-level hidden states $h_1, h_2, \dots, h_T \in \mathbb{R}^d$, where $T$ is the number of frames and $d$ the hidden dimension. They are mean-pooled over time to a single fixed-dimensional embedding $\bar{h} = \frac{1}{T} \sum_{t=1}^{T} h_t$. A linear layer (noted as CLS Head in Fig.~\ref{fig:cms}(a)) then maps $\bar{h}$ to logits over the two classes, from which \texttt{real} and \texttt{fake} probabilities are obtained.

We use two front-end encoders. The first is XLS-R~\cite{babu2022xlsr}, a self-supervised cross-lingual speech encoder; the second is the Whisper encoder~\cite{radford2023whisper}, trained with weak supervision for automatic speech recognition. Their difference is deliberate: XLS-R is speech-native, whereas the Whisper encoder is trained on audio recordings and optimised for recognition rather than for speech representation alone. 

\subsection{ALLM-based CM}

\begin{figure}
    \centering
    \includegraphics[width=1\linewidth]{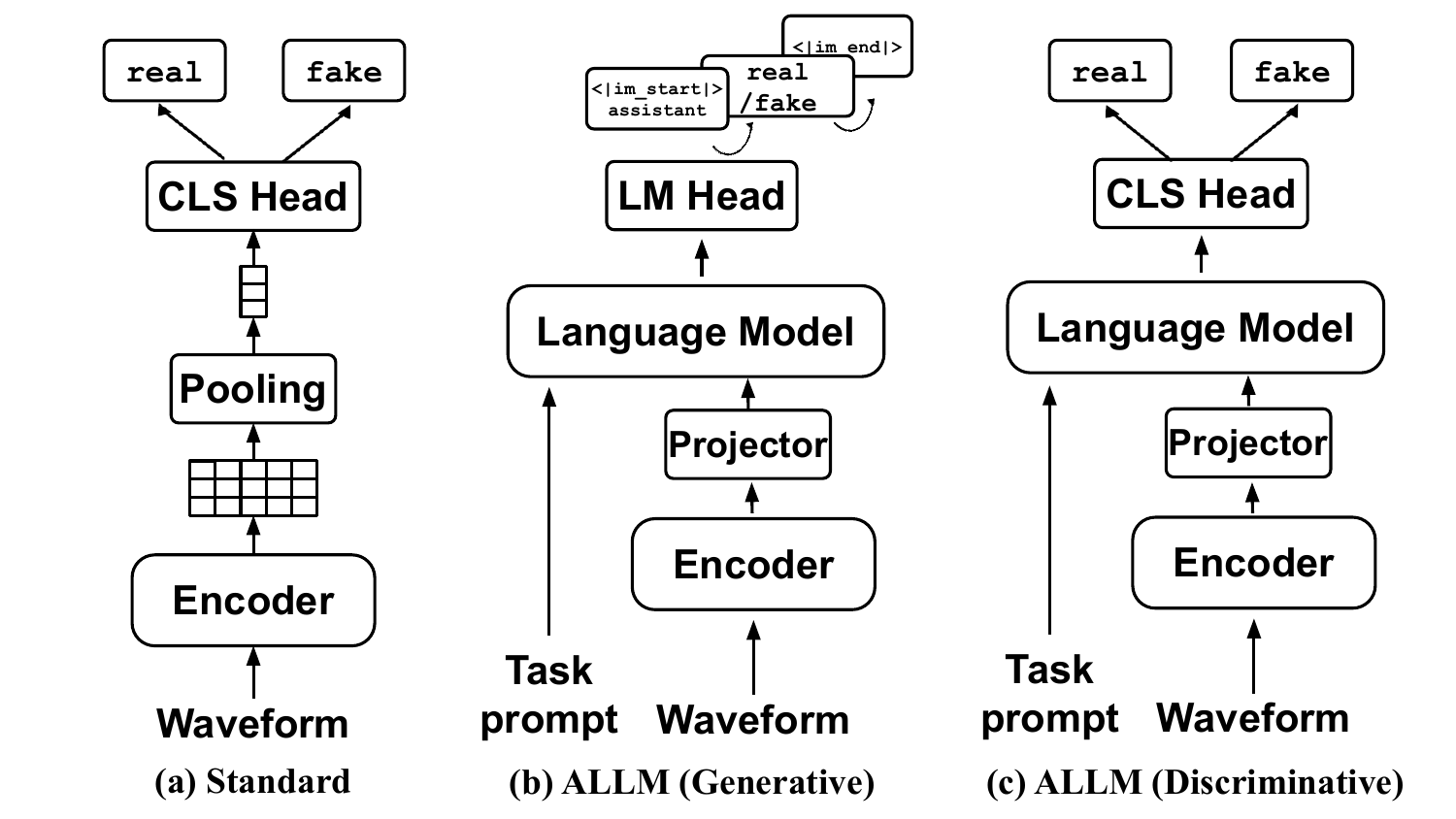}
    \caption{Comparison between different CM architectures.}
    \label{fig:cms}
\end{figure}

An ALLM-based CM (Fig.~\ref{fig:cms}(b), (c)) receives the raw waveform together with a fixed task prompt as input. An audio encoder extracts an embedding sequence from the waveform; a linear projector then maps it to the same dimension as the LLM embedding. The text prompt (shown below) consists of a system message that defines the task role and a user instruction that asks the model to perform classification.  This prompt is tokenised and embedded by the LLM's embedding layer, and the projected audio embeddings (\texttt{<audio>} in the example) are inserted into this sequence before it is passed to the LLM.

\begin{tcolorbox}[
    colback=gray!1,
    colframe=gray!35!black,
    title={LLM Prompt Structure},
    fonttitle=\small\bfseries,
    fontupper=\small,
    arc=1mm,
    boxrule=0.8pt,
    halign=justify
]

[SYSTEM PROMPT]:
\begin{itemize}
    \item Role: \texttt{<|im\_start|>}system
    \item Content: You are a speech forensics expert. You understand prosody, phonetics, voice quality, and speaker naturalness.\texttt{<|im\_end|>}
\end{itemize}
[USER PROMPT]
\begin{itemize}
    \item Role: \texttt{<|im\_start|>}user
    \item Content: Audio 1: \texttt{<audio>} Listen carefully to this recording. Consider whether the voice sounds natural, whether prosody matches the spoken content, and whether the speaker sounds consistent. Classify as either real or fake.\texttt{<|im\_end|>}
\end{itemize}

[ASSISTANT PROMPT]
\begin{itemize}
    \item Role: \texttt{<|im\_start|>}assistant
\end{itemize}

\end{tcolorbox}

We build both ALLM-based CMs with LoRA~\cite{hu2022lora} applied to their audio encoder and transformer-based backbone; the audio embeddings inserted into the prompt are those extracted by a Whisper encoder. Following~\cite{pang2026on}, we read a decision out of the adapted LLMs in two ways:

\paragraph{Generative} 
The LLM operates autoregressively. The two classes are verbalised as the single tokens \texttt{real} and \texttt{fake}, generated at the first response step after the assistant prompt. Their respective logits are extracted from the language modelling head (LM Head in Fig.~\ref{fig:cms}(b)), and a binary softmax is applied to these values to yield a detection score~\cite{wang2025speechllm, gu2025allm4add}.

\paragraph{Discriminative} 
The LLM processes the same prompt sequence, but the language modelling head is replaced with a binary classification head. The logits are therefore read directly from this head to determine the final decision, rather than relying on a generated text token~\cite{pang2026on}.

\section{Experimental setting}
\label{sec:exp_setting}

\subsection{Data and protocols}
We use the CompSpoof V2 database~\cite{zhang2026esdd2overview}, which mixes a speech component and an environmental event component,
each independently real or fake (Table~\ref{tab:classes}).
During CM training, the all-real mixture
\texttt{bonafide\_bonafide} is treated as the \texttt{real} class\footnote{We exclude the unmixed \texttt{Original} class to avoid short-cut learning towards the mixing operation rather than a true synthesis artefact.}, and any condition containing a fake component
(\texttt{bonafide\_spoof}, \texttt{spoof\_bonafide}, \texttt{spoof\_spoof}) is treated as the \texttt{fake} class.

\begin{table}[t] 
\centering 
\caption{Class definitions, labels, and file
counts for model training.} 
\label{tab:classes} 
\resizebox{\columnwidth}{!}{
\begin{tabular}{lcccc} 
\toprule 
\textbf{Class} & \textbf{Speech} & \textbf{Event sound} & \textbf{Training label} & \textbf{\# files} \\ 
\midrule 
\texttt{bonafide\_bonafide} & real & real & \texttt{real}     & 25,189 \\ 
\texttt{bonafide\_spoof}    & real & fake & \texttt{fake}     & 50,361 \\ 
\texttt{spoof\_bonafide}    & fake & real & \texttt{fake}     & 21,759 \\ 
\texttt{spoof\_spoof}       & fake & fake & \texttt{fake}     & 29,413 \\ 
\bottomrule 
\end{tabular} 
}
\end{table}

\begin{table}[t] 
\centering 
\caption{Trial pairings and file counts for EER evaluation on the test set. The target class is always \texttt{bonafide\_bonafide}; the same trained CM is scored once per condition; no CM is retrained.} 
\label{tab:eer_pairings} 
\resizebox{\columnwidth}{!}{%
\begin{tabular}{llclc} 
\toprule 
\textbf{EER condition} & \textbf{Target Class} & \textbf{\# Target} & \textbf{Non-Target Class} & \textbf{\# Non-Target} \\ 
\midrule 
Fake event & \multirow{3}{*}{\texttt{bonafide\_bonafide}} & \multirow{3}{*}{3,635} & \texttt{bonafide\_spoof}    & 7,672 \\ 
Fake speech          &                                              &                        & \texttt{spoof\_bonafide}    & 2,987 \\ 
Both fake &                                              &                        & \texttt{spoof\_spoof}       & 5,894 \\ 
\bottomrule 
\end{tabular}%
}
\end{table}

\subsection{Metric}
We report results according to the equal error rate (EER) on the CompSpoof V2 test set. For each non-target class (\texttt{bonafide\_spoof}, \texttt{spoof\_bonafide}, \texttt{spoof\_spoof}), we pool the detector's output scores on the genuine reference \texttt{bonafide\_bonafide} against that one class and compute the EER. A lower EER indicates that the detector separates genuine mixtures from that particular fake condition more reliably. 

\subsection{Implementation details}

All four CMs operate on 16 kHz audio and are optimised using cross-entropy loss with AdamW~\cite{loshchilov2018decoupled} and class-balanced sampling. For the generative ALLM CM, target decisions are verbalised as the single tokens \texttt{real} or \texttt{fake}; cross-entropy loss is computed only at the target response position over the full LM vocabulary logits. For the remaining three discriminative CMs, parameters are optimised using cross-entropy loss between the CLS-head output logits and the binary ground-truth labels.

Every CM uses a peak learning rate of $4 \times 10^{-5}$ with 200 warm-up steps, a batch size of 8, validation every 200 steps,  and a 30,000-step optimiser limit (up to five epochs), with the best checkpoint selected based on the lowest validation loss. The two non-ALLM CMs share an identical recipe that all system components are fine-tuned end-to-end. The two ALLM-based CMs are built with Qwen2-Audio-7B-Instruct~\cite{chu2024qwen2audio} and adapted with LoRA ($\text{rank}=16$ and $\alpha=32$). For the discriminative setting, a linear classification head over the pooled hidden state is trained jointly with the LoRA parameters.

All models and experiments are run on a single NVIDIA H100 GPU. Performance is estimated from the average of three independent runs each with a different random seed.

\subsection{Sub-band probing}

To localise where each detector's evidence concentrates~(RQ2), we adopt the band-stop probing protocol of~\cite{wang2022investigating}. The 0--8~kHz range is partitioned into six bands
(0--0.8, 0.8--2.4, 2.4--4, 4--5.6, 5.6--7.2, and 7.2--8~kHz); one band at a time is suppressed on the raw waveform with an 8th-order Butterworth filter applied in zero-phase form to avoid phase distortion. The four middle bands use standard band-stop filters; the two edge bands are realised as one-sided filters. Filtering is applied only at inference: the same trained CM is reused for every condition with no retraining to compute the per-condition EER.

\section{RQ1 Results and analysis}
\label{sec:rq1}

In this section, we focus on our first research question: when speech and an event are mixed, which component does a detector catch more readily? EER results are reported in Section~\ref{sec:component_eer}, an additional analysis on the fake speech component is presented in Section~\ref{sec:speech_decompose}.

\begin{table}[t]
  \centering
  \caption{EER~(\%) on the test set, reported as mean~{\scriptsize\color{black!55}(standard deviation)} over three random seeds. Cells are colour-coded in grayscale by row, with darker and lighter shades indicating higher and lower mean EER.}
  \label{tab:three_eers}
  \renewcommand{\arraystretch}{1.25}
  \setlength{\tabcolsep}{6pt}
  \resizebox{\columnwidth}{!}{%
  \begin{tabular}{lccc}
    \toprule
\textbf{Detector} & \textbf{Fake event EER} & \textbf{Fake speech EER} & \textbf{Both fake EER} \\
    \midrule
    XLS-R    & \cellcolor[rgb]{1.00, 1.00, 1.00} 9.50 {\scriptsize\color{black!55} (0.22)} & \cellcolor[rgb]{0.59, 0.59, 0.59} 21.21 {\scriptsize\color{black!55} (0.68)} & \cellcolor[rgb]{0.90, 0.90, 0.90} 13.96 {\scriptsize\color{black!55} (0.77)} \\
    Whisper    & \cellcolor[rgb]{1.00, 1.00, 1.00} 8.29 {\scriptsize\color{black!55} (0.62)} & \cellcolor[rgb]{0.59, 0.59, 0.59} 23.15 {\scriptsize\color{black!55} (0.59)} & \cellcolor[rgb]{0.90, 0.90, 0.90} 13.89 {\scriptsize\color{black!55} (0.45)} \\
    Qwen2-Audio~(disc)    & \cellcolor[rgb]{1.00, 1.00, 1.00} 13.38 {\scriptsize\color{black!55} (0.47)} & \cellcolor[rgb]{0.59, 0.59, 0.59} 23.74 {\scriptsize\color{black!55} (0.36)} & \cellcolor[rgb]{0.89, 0.89, 0.89} 17.28 {\scriptsize\color{black!55} (0.67)} \\
    Qwen2-Audio~(gen)    & \cellcolor[rgb]{1.00, 1.00, 1.00} 12.91 {\scriptsize\color{black!55} (1.03)} & \cellcolor[rgb]{0.59, 0.59, 0.59} 23.25 {\scriptsize\color{black!55} (0.60)} & \cellcolor[rgb]{0.90, 0.90, 0.90} 16.89 {\scriptsize\color{black!55} (0.45)} \\
    \bottomrule
  \end{tabular}%
  }
\end{table}

\subsection{Component sensitivity}
\label{sec:component_eer}

Table~\ref{tab:three_eers} reports EER results for the four CM systems under the three evaluated conditions. Results when only one component is fake (columns 2 and 3) show that synthetic speech is harder to separate from a genuine mixture than a synthetic event. The lowest EER for each CM occurs when only the event is fake: 9.50\% for XLS-R, 8.29\% for Whisper, 13.40\% for Qwen2-Audio~(disc), and 12.94\% for Qwen2-Audio~(gen). Conversely, the highest EER for every CM occurs when only the speech is fake: 21.21\%, 23.15\%, 23.73\%, and 23.25\%, respectively. Though evaluated under a different metric, this finding is consistent with the ESDD2 challenge systems, for which the environmental sound EER is frequently lower than the speech EER~\cite{zhang2026esdd2overview}.

Intuitively the both fake condition (column 4) should be the hardest, since the CM could be confused by either component. It is, however, easier to flag than the fake speech condition. Three hypotheses may account for this observation: (a) the CM may be biased towards artefacts introduced by synthetic event sound, which are absent in the fake speech condition; (b) a genuine event may mask the speech artefact; (c) the fake speech may simply be harder to detect than the fake events. We investigate these hypotheses in the following section.

\subsection{Decomposed EER}
\label{sec:speech_decompose}

Table~\ref{tab:speech_eers} shows EER results decomposed by fake speech sources in the CompSpoof V2 test set, including SoulX-Podcast~\cite{xie2025soulxpodcast}, MLAAD~\cite{muller2024mlaad}, and ASVspoof 5~\cite{wang24_asvspoof5}. 

EERs for MLAAD and ASVspoof 5 shift by at most a few percentage points between conditions. This indicates that the presence of a fake or a real event sound does not substantially alter the CM's access to speech artefacts. It contradicts the masking hypothesis (b) of the previous section. Furthermore, the two sources appear both in the training and test partitions~\cite{zhang2026esdd2}, and when such seen speech artefacts are accessible, the CMs do not inherently rely on fake event cues regardless of whether they are present or not, which contradicts the bias hypothesis (a).

\begin{table}[t]
  \centering
  \caption{EER~(\%) on the test set decomposed by spoofed-speech source, reported as mean~{\scriptsize\color{black!55}(standard deviation)} over three random seeds. Cells are colour-coded in grayscale by row, with darker and lighter shades indicating higher and lower mean EER.}
  \label{tab:speech_eers}
  \renewcommand{\arraystretch}{1.25}
  \setlength{\tabcolsep}{5pt}
  \begin{tabularx}{\columnwidth}{l *{6}{>{\centering\arraybackslash}X}}
\toprule
    & \multicolumn{3}{c}{\textbf{Fake speech EER}}
    & \multicolumn{3}{c}{\textbf{Both fake EER}} \\
    \cmidrule(lr){2-4} \cmidrule(lr){5-7}
    \textbf{Detector}
      & \textbf{SoulX} & \textbf{MLAAD~} & \textbf{~ASV5}
      & \textbf{SoulX} & \textbf{MLAAD~} & \textbf{~ASV5} \\
\midrule
    XLS-R  & \cellcolor[rgb]{0.55, 0.55, 0.55} 36.70 {\scriptsize\color{black!55} (0.96)} & \cellcolor[rgb]{0.99, 0.99, 0.99} 4.61 {\scriptsize\color{black!55} (0.36)} & \cellcolor[rgb]{1.00, 1.00, 1.00} 4.08 {\scriptsize\color{black!55} (0.30)} & \cellcolor[rgb]{0.83, 0.83, 0.83} 16.34 {\scriptsize\color{black!55} (0.93)} & \cellcolor[rgb]{0.99, 0.99, 0.99} 4.85 {\scriptsize\color{black!55} (0.39)} & \cellcolor[rgb]{0.98, 0.98, 0.98} 5.50 {\scriptsize\color{black!55} (1.78)} \\
    Whisper     & \cellcolor[rgb]{0.55, 0.55, 0.55} 41.43 {\scriptsize\color{black!55} (0.91)} & \cellcolor[rgb]{0.98, 0.98, 0.98} 4.21 {\scriptsize\color{black!55} (0.49)} & \cellcolor[rgb]{1.00, 1.00, 1.00} 2.86 {\scriptsize\color{black!55} (0.58)} & \cellcolor[rgb]{0.83, 0.83, 0.83} 17.15 {\scriptsize\color{black!55} (0.59)} & \cellcolor[rgb]{0.99, 0.99, 0.99} 3.76 {\scriptsize\color{black!55} (0.37)} & \cellcolor[rgb]{0.99, 0.99, 0.99} 3.68 {\scriptsize\color{black!55} (0.27)} \\
    Qwen2-Audio~(disc) & \cellcolor[rgb]{0.55, 0.55, 0.55} 49.52 {\scriptsize\color{black!55} (2.85)} & \cellcolor[rgb]{0.97, 0.97, 0.97} 9.03 {\scriptsize\color{black!55} (0.68)} & \cellcolor[rgb]{1.00, 1.00, 1.00} 6.81 {\scriptsize\color{black!55} (0.32)} & \cellcolor[rgb]{0.82, 0.82, 0.82} 20.42 {\scriptsize\color{black!55} (0.87)} & \cellcolor[rgb]{0.97, 0.97, 0.97} 8.49 {\scriptsize\color{black!55} (0.69)} & \cellcolor[rgb]{0.99, 0.99, 0.99} 7.97 {\scriptsize\color{black!55} (0.05)} \\
    Qwen2-Audio~(gen) & \cellcolor[rgb]{0.55, 0.55, 0.55} 41.52 {\scriptsize\color{black!55} (0.52)} & \cellcolor[rgb]{0.96, 0.96, 0.96} 9.56 {\scriptsize\color{black!55} (0.34)} & \cellcolor[rgb]{1.00, 1.00, 1.00} 6.65 {\scriptsize\color{black!55} (0.79)} & \cellcolor[rgb]{0.83, 0.83, 0.83} 19.64 {\scriptsize\color{black!55} (0.54)} & \cellcolor[rgb]{0.98, 0.98, 0.98} 8.07 {\scriptsize\color{black!55} (0.43)} & \cellcolor[rgb]{1.00, 1.00, 1.00} 6.89 {\scriptsize\color{black!55} (1.39)} \\
    \bottomrule
  \end{tabularx}
\end{table}

\begin{table*}[t]
  \centering
  \caption{EER~(\%) on the test set under sub-band probing, reported as mean~{\scriptsize\color{black!55}(standard deviation)} over three random seeds. The Original column applies no band-stop filter and closely reproduces Table~\ref{tab:three_eers}. Cells are colour-coded in grayscale by row, with darker and lighter shades indicating higher and lower mean EER.}
  \label{tab:bandstop}

  \renewcommand{\arraystretch}{1.25}

  \smallskip
  \noindent\textbf{(a) Fake event EER}

  \smallskip
  \setlength{\tabcolsep}{5pt}
  \begin{tabularx}{\textwidth}{l *{7}{>{\centering\arraybackslash}X}}
    \toprule
    \textbf{Detector}
      & \textbf{Original}
      & \textbf{0--0.8 kHz}
      & \textbf{0.8--2.4 kHz}
      & \textbf{2.4--4 kHz}
      & \textbf{4--5.6 kHz}
      & \textbf{5.6--7.2 kHz}
      & \textbf{7.2--8 kHz} \\
    \midrule
    XLS-R & \cellcolor[rgb]{0.99, 0.99, 0.99} 9.50 {\scriptsize\color{black!55} (0.22)} & \cellcolor[rgb]{0.95, 0.95, 0.95} 12.78 {\scriptsize\color{black!55} (0.23)} & \cellcolor[rgb]{0.97, 0.97, 0.97} 11.39 {\scriptsize\color{black!55} (0.21)} & \cellcolor[rgb]{1.00, 1.00, 1.00} 9.00 {\scriptsize\color{black!55} (0.16)} & \cellcolor[rgb]{0.98, 0.98, 0.98} 10.43 {\scriptsize\color{black!55} (1.64)} & \cellcolor[rgb]{0.94, 0.94, 0.94} 13.15 {\scriptsize\color{black!55} (0.80)} & \cellcolor[rgb]{0.59, 0.59, 0.59} 26.15 {\scriptsize\color{black!55} (2.47)} \\
    Whisper & \cellcolor[rgb]{1.00, 1.00, 1.00} 8.29 {\scriptsize\color{black!55} (0.62)} & \cellcolor[rgb]{0.84, 0.84, 0.84} 18.30 {\scriptsize\color{black!55} (1.25)} & \cellcolor[rgb]{0.65, 0.65, 0.65} 25.34 {\scriptsize\color{black!55} (2.41)} & \cellcolor[rgb]{0.96, 0.96, 0.96} 11.25 {\scriptsize\color{black!55} (1.60)} & \cellcolor[rgb]{0.98, 0.98, 0.98} 10.26 {\scriptsize\color{black!55} (0.37)} & \cellcolor[rgb]{0.94, 0.94, 0.94} 13.44 {\scriptsize\color{black!55} (1.31)} & \cellcolor[rgb]{0.59, 0.59, 0.59} 27.23 {\scriptsize\color{black!55} (1.30)} \\
    Qwen2-Audio~(disc) & \cellcolor[rgb]{0.90, 0.90, 0.90} 13.38 {\scriptsize\color{black!55} (0.47)} & \cellcolor[rgb]{0.85, 0.85, 0.85} 14.61 {\scriptsize\color{black!55} (1.01)} & \cellcolor[rgb]{0.97, 0.97, 0.97} 11.31 {\scriptsize\color{black!55} (0.35)} & \cellcolor[rgb]{0.93, 0.93, 0.93} 12.40 {\scriptsize\color{black!55} (0.47)} & \cellcolor[rgb]{0.95, 0.95, 0.95} 11.91 {\scriptsize\color{black!55} (0.54)} & \cellcolor[rgb]{1.00, 1.00, 1.00} 9.76 {\scriptsize\color{black!55} (0.51)} & \cellcolor[rgb]{0.59, 0.59, 0.59} 19.47 {\scriptsize\color{black!55} (0.36)} \\
    Qwen2-Audio~(gen) & \cellcolor[rgb]{0.94, 0.94, 0.94} 12.91 {\scriptsize\color{black!55} (1.03)} & \cellcolor[rgb]{0.92, 0.92, 0.92} 13.08 {\scriptsize\color{black!55} (0.30)} & \cellcolor[rgb]{0.98, 0.98, 0.98} 11.11 {\scriptsize\color{black!55} (0.63)} & \cellcolor[rgb]{0.96, 0.96, 0.96} 11.62 {\scriptsize\color{black!55} (0.74)} & \cellcolor[rgb]{0.97, 0.97, 0.97} 11.33 {\scriptsize\color{black!55} (0.80)} & \cellcolor[rgb]{1.00, 1.00, 1.00} 9.97 {\scriptsize\color{black!55} (0.38)} & \cellcolor[rgb]{0.59, 0.59, 0.59} 20.71 {\scriptsize\color{black!55} (0.68)} \\
    \bottomrule
  \end{tabularx}

  \bigskip
  \noindent\textbf{(b) Fake speech EER}

  \smallskip
  \setlength{\tabcolsep}{5pt}
  \begin{tabularx}{\textwidth}{l *{7}{>{\centering\arraybackslash}X}}
    \toprule
    \textbf{Detector}
      & \textbf{Original}
      & \textbf{0--0.8 kHz}
      & \textbf{0.8--2.4 kHz}
      & \textbf{2.4--4 kHz}
      & \textbf{4--5.6 kHz}
      & \textbf{5.6--7.2 kHz}
      & \textbf{7.2--8 kHz} \\
    \midrule
    XLS-R & \cellcolor[rgb]{1.00, 1.00, 1.00} 21.21 {\scriptsize\color{black!55} (0.68)} & \cellcolor[rgb]{0.59, 0.59, 0.59} 32.53 {\scriptsize\color{black!55} (0.46)} & \cellcolor[rgb]{0.95, 0.95, 0.95} 23.67 {\scriptsize\color{black!55} (0.92)} & \cellcolor[rgb]{0.92, 0.92, 0.92} 24.81 {\scriptsize\color{black!55} (0.56)} & \cellcolor[rgb]{0.96, 0.96, 0.96} 23.26 {\scriptsize\color{black!55} (1.02)} & \cellcolor[rgb]{0.92, 0.92, 0.92} 24.66 {\scriptsize\color{black!55} (0.60)} & \cellcolor[rgb]{0.98, 0.98, 0.98} 22.37 {\scriptsize\color{black!55} (0.96)} \\
    Whisper & \cellcolor[rgb]{0.99, 0.99, 0.99} 23.15 {\scriptsize\color{black!55} (0.59)} & \cellcolor[rgb]{0.77, 0.77, 0.77} 26.68 {\scriptsize\color{black!55} (0.62)} & \cellcolor[rgb]{0.95, 0.95, 0.95} 23.95 {\scriptsize\color{black!55} (1.55)} & \cellcolor[rgb]{0.82, 0.82, 0.82} 26.06 {\scriptsize\color{black!55} (0.99)} & \cellcolor[rgb]{1.00, 1.00, 1.00} 22.86 {\scriptsize\color{black!55} (1.14)} & \cellcolor[rgb]{0.90, 0.90, 0.90} 24.93 {\scriptsize\color{black!55} (0.97)} & \cellcolor[rgb]{0.59, 0.59, 0.59} 28.43 {\scriptsize\color{black!55} (2.45)} \\
    Qwen2-Audio~(disc) & \cellcolor[rgb]{0.84, 0.84, 0.84} 23.74 {\scriptsize\color{black!55} (0.36)} & \cellcolor[rgb]{0.59, 0.59, 0.59} 25.82 {\scriptsize\color{black!55} (0.47)} & \cellcolor[rgb]{0.73, 0.73, 0.73} 25.09 {\scriptsize\color{black!55} (0.82)} & \cellcolor[rgb]{0.81, 0.81, 0.81} 24.08 {\scriptsize\color{black!55} (0.51)} & \cellcolor[rgb]{0.88, 0.88, 0.88} 23.33 {\scriptsize\color{black!55} (0.38)} & \cellcolor[rgb]{0.95, 0.95, 0.95} 22.66 {\scriptsize\color{black!55} (0.56)} & \cellcolor[rgb]{1.00, 1.00, 1.00} 21.39 {\scriptsize\color{black!55} (0.48)} \\
    Qwen2-Audio~(gen) & \cellcolor[rgb]{0.86, 0.86, 0.86} 23.25 {\scriptsize\color{black!55} (0.60)} & \cellcolor[rgb]{0.59, 0.59, 0.59} 25.63 {\scriptsize\color{black!55} (0.41)} & \cellcolor[rgb]{0.72, 0.72, 0.72} 24.67 {\scriptsize\color{black!55} (0.39)} & \cellcolor[rgb]{0.84, 0.84, 0.84} 23.48 {\scriptsize\color{black!55} (0.36)} & \cellcolor[rgb]{0.89, 0.89, 0.89} 22.87 {\scriptsize\color{black!55} (0.56)} & \cellcolor[rgb]{0.97, 0.97, 0.97} 21.79 {\scriptsize\color{black!55} (0.24)} & \cellcolor[rgb]{1.00, 1.00, 1.00} 21.32 {\scriptsize\color{black!55} (0.33)} \\
    \bottomrule
  \end{tabularx}

\end{table*}

The unseen SoulX-Podcast source behaves differently. Its EER falls when the event is also fake (e.g., Fake speech EER from 36.70–49.52\% to Both fake EER 16.34–20.42\%). Given the observation that event sound has little impact on the speech artefact, it is likely that the unseen fake speech is inherently harder to detect (hypothesis (c)). The added event artefact supplies an additional, easy detection cue that partially compensates for the generalisation gap on the speech side.

\section{RQ2  Results and analysis}
\label{sec:rq2}

In this section, we focus on our second research question: which frequency sub-bands carry the detector's evidence. EER results are reported in Section~\ref{sec:subband}, the corresponding score distributions are reported in Section~\ref{sec:score_distribution}.

\subsection{Where does the evidence live?}
\label{sec:subband}

Table~\ref{tab:bandstop} reports EER when band-stop filtering is applied to the input waveform for the fake event (Table~\ref{tab:bandstop}(a)) and fake speech (Table~\ref{tab:bandstop}(b)) conditions. 

When event is fake, the top band carries the evidence for every CM. As shown in Table~\ref{tab:bandstop}(a), muting 7.2–8 kHz is the single most damaging perturbation across frequency axis: EER rises from 9.50\% to 26.15\% for XLS-R, from 8.29\% to 27.23\% for Whisper, from 13.38\% to 19.47\% for Qwen2-Audio (disc), and from 12.91\% to 20.71\% for Qwen2-Audio (gen). The cue that exposes a synthetic event therefore concentrates more at the upper edge of the representable spectrum.

We notice that for the Whisper classifier the 0.8--2.4~kHz band is nearly as damaging (25.34\%) as the top band. The two Qwen2-Audio systems, although they too use the Whisper encoder, do not inherit this secondary sensitivity. Instead, masking any of the four intermediate bands (0.8--7.2~kHz) leaves their EER below the original value. These sub-bands appear to act as mild distractions whose removal can marginally benefit deepfake detection.

When speech is fake, only the speech-native XLS-R front-end relies wholly on the vocal band. As shown in Table~\ref{tab:bandstop}(b), muting the lowest band (0–0.8 kHz) is the most damaging perturbation for XLS-R (21.21\% to 32.53\%). The same holds for the two ALLM-based CMs, but only by a small margin: Qwen2-Audio (disc) from 23.74\% to 25.82\%, Qwen2-Audio (gen) from 23.25\% to 25.63\%. For the Whisper classifier, muting the top band remains the most damaging perturbation (28.43\%). Three CMs that adopt Whisper as a front-end (bottom three rows in Table~\ref{tab:bandstop}(b)) appear to spread their attention more evenly across the frequency axis when detecting artefacts.

Interestingly, for both ALLM-based CMs, performance on the fake speech condition (Table~\ref{tab:bandstop}(b)) is at its best when the top band is removed: the discriminative model improves from 23.74\% to 21.39\% and the generative model from 23.25\% to 21.32\%. This contradicts the observation from Table~\ref{tab:bandstop}(a) and the top band appears to provide a misleading cue to the ALLM when the question concerns fake speech alone. We analyse this behaviour in the next section.

\subsection{Score distribution}
\label{sec:score_distribution}

\begin{figure}
    \centering
    \includegraphics[width=0.99\linewidth]{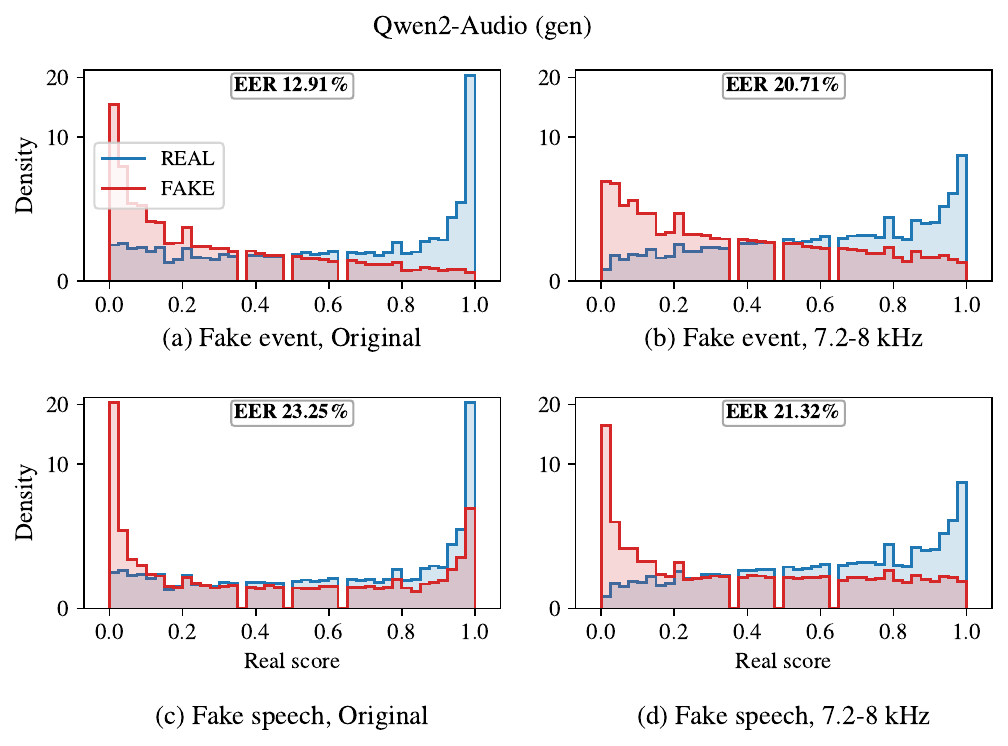}
    \caption{Score distributions of the Qwen2-Audio (gen) system under no filtering (first column) and with 7.2–8 kHz muted (second column). Top row: fake event condition. Bottom row: fake speech condition. }
    \label{fig:score_distribution}
\end{figure}

Fig.~\ref{fig:score_distribution} shows score distributions for the two conditions reported in Table~\ref{tab:bandstop}. The Qwen2-Audio (gen) system and the top band (7.2–8 kHz) EER are chosen as representative for easier visualisation.

Compared to the original distribution (Fig.~\ref{fig:score_distribution}(a)), removing the top band in the fake event condition (Fig.~\ref{fig:score_distribution}(b)) reduces the share of density at both edges of the x-axis. They represent the CM's high-confidence decisions for both \texttt{real} and the \texttt{fake} class. This shift indicates that the detector is less confident once the top band is gone, and such reduced confidence results in a drop of near 8\% in EER (12.91\% to 20.71\%).

The fake speech condition of the bottom row exhibits a different behaviour. The \texttt{real} class shifts in the same way as in Fig.~\ref{fig:score_distribution}(b), since the two conditions share the same target class and hence an identical real score distribution. The \texttt{fake} class in Fig.~\ref{fig:score_distribution}(d), however, moves the other way: though the highly confident correct predictions (left edge) reduce only marginally, the tail of highly confident incorrect predictions (right edge) is substantially reduced compared to the unfiltered distribution in Fig.~\ref{fig:score_distribution}(c). Such reduction drives the EER down from 23.25\% to 21.32\%.

Fig.~\ref{fig:score_distribution} shows that muting the top band affects the two conditions differently. For fake events, this removal reduces CM confidence across both classes; for fake speech, by contrast, it primarily reduces the CM's highly confident incorrect classifications. This difference suggests that the top band mainly captures event cues, which impacts the CMs more heavily when the event is fake rather than the speech.

\section{Discussion and Conclusion}
\label{sec:conclusion}

We evaluated deepfake detectors using mixtures where the speech and the background event could independently be fake. Our initial finding shows that all four CMs detect fake events far more accurately than fake speech. Though this may suggest that identifying a fake event is inherently easier than detecting fake speech, upon a detailed, decomposed analysis on each fake speech source, we find that such performance gap is largely a generalisation failure towards an unseen podcast-style speech generator~\cite{xie2025soulxpodcast}. This observation is consistent with other literature~\cite{mahapatra25can,zhao2024emofake} that expressive and emotional synthetic speech degrades current CMs whose training data contains mostly neutral, reading speech.

Our four CMs vary two factors by design to allow controlled analysis of the influence of either the representation extractor or the backbone classifier. We find that the encoder is the dominant reason for the difference in detector behaviour. Only the speech-native XLS-R localises its speech-spoof evidence fully in the low vocal band (Table~\ref{tab:bandstop}(a)); the Whisper-based systems read the upper edge regardless of which component is fake
(Table~\ref{tab:bandstop}(b) and Fig.~\ref{fig:score_distribution}). 

The generative and discriminative settings of the same ALLM backbone are nearly indistinguishable by EER (Tables~\ref{tab:three_eers}--\ref{tab:bandstop}),
yet a token-level analysis reveals an internal behaviour difference. We collect the tokens each ALLM outputs for its decision across a randomly selected test subset. For the generative ALLM, the highest-logit tokens are relevant to the task: \texttt{Real}, \texttt{natural}, and
\begin{CJK}{UTF8}{gbsn}真人\end{CJK} (lit.\ real human) for the \texttt{real} class;
and \texttt{Fake}, \texttt{phony}, and \begin{CJK}{UTF8}{gbsn}假\end{CJK}
(lit.\ fake) for the \texttt{fake} class. For the discriminative ALLM, task-relevant
tokens (\texttt{Real}, \begin{CJK}{UTF8}{gbsn}真实\end{CJK}
(lit.\ genuine)) are accompanied by unrelated words such as \texttt{The},
\texttt{It}, and \texttt{Sleep}, which implies that the classification head draws on a less
semantically coherent set of representations.

Future work includes a more detailed investigation into the representations extracted and processed by ALLMs. Analysis can also be extended from environmental sound to more challenging cases such as music~\cite{zang2024singfake,rahman2024sonics}. The CM architectures used in this work adopt a less complex module design to benefit controlled comparison and analysis, though this is not necessarily optimal for performance. It will be of interest to investigate more complex architectures in future work.

\section{Acknowledgments}
This study is supported by JSPS KAKENHI Grant (25K24398). This study was carried out on TSUBAME4.0 supercomputer at the Institute of Science Tokyo.

Generative AI tools were used to assist in generating scripts for model training, result analysis, and visualisation. They were also used to polish the language of the paper.

\balance
\bibliographystyle{IEEEtran}
\bibliography{ref}

\end{document}